# Direct assessment of the proton affinity of individual surface hydroxyls with non-contact atomic force microscopy

Margareta Wagner[1, 2], Bernd Meyer[3], Martin Setvin[1], Michael Schmid[1], and Ulrike Diebold[1]

[1]Institute of Applied Physics, TU Wien, Wiedner Hauptstrasse 8-10/134,
A-1040 Vienna, Austria
[2]Central European Institute of Technology (CEITEC), Brno University of Technology,
Purkyňova 123, 612 00 Brno, Czech Republic
[3]Interdisciplinary Center for Molecular Materials and Computer-Chemistry-Center,
Friedrich-Alexander-Universität Erlangen-Nürnberg, Nägelsbachstrasse 25,
91052 Erlangen, Germany

**The state of protonation/deprotonation of surfaces has far-ranging implications in all areas of chemistry: from acid-base catalysis[1] and the electro- and photocatalytic splitting of water[2], to the behavior of minerals[3] and biochemistry[4]. The acidity of a molecule or a surface site is described by its proton affinity (PA) and p$K_a$ value (the negative logarithm of the equilibrium constant of the proton transfer reaction in solution). For solids, in contrast to molecules, the acidity of individual sites is difficult to assess. For mineral surfaces such as oxides they are estimated by semi-empirical concepts such as bond-order valence sums[5], and also increasingly modeled with first-principles molecular dynamics simulations[6,7]. Currently such predictions cannot be tested – the experimental measures used for comparison are typically average quantities integrated over the whole surface or, in some cases, individual crystal facets[8], such as the point of zero charge (pzc)[9].**

**Here we assess individual hydroxyls on $In_2O_3(111)$, a model oxide with four different types of surface oxygen atoms, and probe the strength of their hydrogen bond with the tip of a non-contact atomic force microscope (AFM). The force curves are in quantitative agreement with density-functional theory (DFT) calculations. By relating the results to known proton affinities and p$K_a$ values of gas-phase molecules, we provide a direct measure of proton affinity distributions at the atomic scale.**

Our work relies on recent developments in high-resolution, non-contact atomic force microscopy (nc-AFM) using the qPlus sensor[10]. Appropriately functionalized tips[11] allow imaging the chemical structure of molecules, evaluating surface structures[12], and following the diffusion of water clusters in a non-perturbative way[13]. Most importantly, one can probe the properties of individual surface atoms[14], such as their chemical nature[15], their charge[16], or their electronegativity[17].

Here we show that an OH-functionalized AFM tip can be used for quantitative insights into the acidity of individual surface OH groups. Since our AFM measurements are conducted in vacuum, we directly assess the proton affinity (PA), the key contribution to the p$K_a$ value. The PA is defined as the enthalpy change upon deprotonating an OH group in the gas phase[18] and the p$K_a$ value is then obtained by adding the Gibbs free energies of solvation (see Method Section). While it is difficult to determine absolute PA values, differences can be measured by gas phase titration with probe molecules with a known PA; in case of a strong base by measuring the proton transfer, for a weak base, where no proton transfer occurs, by measuring adsorption enthalpies[19]. With the AFM tip we work in the weak basicity limit: an attached OH group acts as the probe molecule; by measuring its interaction strength with a surface OH we can attribute local PAs to individual sites.

A particularly good test system for evaluating the PA of an oxide is the $In_2O_3(111)$ surface, as it offers a multitude of well-defined surface sites, and has promising practical applications for the photocatalytic conversion[20] of $CO_2$ and in electro-[21] and heterogeneous catalysis[22,23]. The material crystallizes in the bixbyite structure (space group Ia-3) with a cubic unit cell of 1.0117 nm. In the bulk In atoms are 6-fold coordinated (In(6c)) and each O(4c) atom has four bonds of unequal lengths, ranging from 2.13 to 2.23 Å.[24]

The (111) surface consists of an O-In-O tri-layer and is stable in its bulk-terminated, (1×1) form (Fig. 1a) with small relaxations[25]. Of the 16 In atoms in the surface unit cell, four remain In(6c), while the 12 In(5c) miss one bond. The twelve O(3c) atoms in the unit cell fall into four categories (α-δ), each type misses a bond of one particular length. The different coordination environments of these surface O atoms are reflected in their electronic structure. Fig. 1b shows the valence band region, which, as is true for most oxides, is mostly of O-2$p$ character. The O(α) atoms are bonded to three In(5c) in a strongly relaxed environment and their partial density of states (PDOS) resembles those of bulk O. The states of the O(β) are highest in energy, and the positions of the valence band states of the O(γ) and of the O(δ) fall in-between. From the distinct positioning of the frontier orbitals, one would expect a clear trend in reactivity. Indeed, the calculated adsorption energies of an isolated H atom (in the legend of Fig. 1b) follow these expectations.

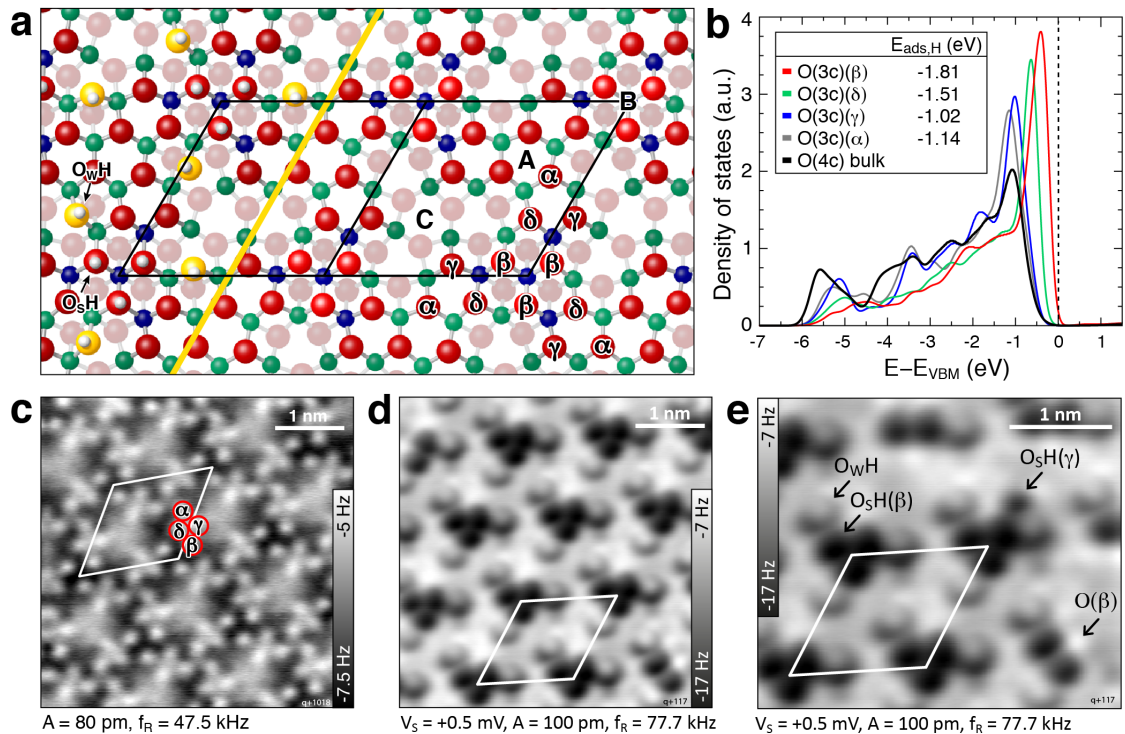

**Fig. 1 The clean and hydroxylated In$_2$O$_3$(111) surface**. (a) Structural model of the relaxed (1×1) surface with the positions of 3-fold symmetry axes labeled A, B, C. Each surface unit cell contains four 6-fold coordinated In(6c) atoms (blue), 12 In(5c) atoms (green), 12 O(3c) and 12 O(4c) atoms (red). The four types of O(3c) atoms are labeled α, β, γ, and δ. At saturation, left of the yellow line, three water molecules adsorb dissociatively at equivalent sites, forming an O$_W$H hydroxyl and a surface hydroxyl at an O(3c)(β). (b) Partial density of states of the various O(3c) atoms, plotted with respect to the valence band maximum (VBM). The calculated adsorption energies E$_{ads}$ of H atoms for each of the O(3c) are listed in the inset. (c) Constant-height nc-AFM image taken with an O-terminated tip in repulsion. (d) Non-contact AFM image of the surface saturated by exposure to water vapor at room temperature. The O$_W$H and O$_S$H(β) sit at the positions indicated in Fig.1b, and are clearly distinguished by their frequency shift. (e) Surface after a proton was transferred from an O(β) to an O(γ) site by tip manipulation.

We have prepared the (111) surface of an In$_2$O$_3$ single crystal in ultrahigh vacuum (UHV), following previously-established preparation procedures[25,26]. The sample was imaged with a low-temperature STM/nc-AFM equipped with a qPlus sensor. The image in Fig. 1c was obtained with an O-terminated tip (prepared by first scanning a TiO$_2$ surface covered with O$_2$, see ExtendedData Fig. 1), and displays the shift in resonance frequency during a constant-height scan. The individual O(3c) are clearly distinguished. Representative AFM images with other tip terminations are shown in ExtendedData Fig. 1.

To hydroxylate the surface, the sample was exposed to water vapor at room temperature. Under these conditions water adsorbs dissociatively[27] with a saturation coverage of 3 molecules/unit cell. Interestingly, only one spot of the large unit cell is active for adsorption: the hydroxyl stemming from the water molecule ($O_WH$) always bridges two In(5c), and the split-off proton forms a surface hydroxyl ($O_SH$) at the neighboring O(β) (sketched at the left side of Fig. 1a). This results in a 'propeller-type' configuration of three equivalent $O_WH$-$O_SH$(β) pairs, as seen in the nc-AFM image in Fig. 1d. The precise location of these hydroxyls was discerned from images taken with different water coverages and tip-sample distances, see ExtendedData Figs. 3 and 4. The hydroxylation of the surface by adsorption of 3 molecules/unit cell leads to only small changes in the electronic structure and the reactivity of the unprotonated surface sites, see ExtendedData Fig. 7.

To construct hydroxyls at other surface O(3c) atoms, we resorted to manipulations with the STM tip (see ExtendedData Fig. 5). Voltage pulsing and/or scanning with >3.5 V desorbs protons from the $O_SH$.[27] The parameters that resulted in the transfer of a proton are close to the threshold for its desorption, and in general, it re-adsorbed on an O(3c) close to the original $O_SH$(β). We could create $O_SH$(γ) (as in Fig. 1e) and $O_SH$(δ), but the O(α) are located farther from the site where the water adsorbs, and formation of an $O_SH$(α) was observed only once during our experiments.

Figure 2a shows short-range force-distance curves taken on the various hydroxyls. These measurements were robust and reproducible; several data sets are shown. The $z$-positions of the minima scales with the vertical position of the hydroxyl's O atom (as reference, $z = 0$ was set at the position of the $O_WH$), and a clear trend is seen for the minima in the force curves.

The force-distance curves were modeled from first principles using DFT (Fig. 2b). Several tips were tried. Most of them were unstable; a tip composed of indium oxide and terminated with an OH group was stable and inert when bringing it close to the surface (see Fig. 2c and also ExtendedData Fig. 8). It is reasonable to assume that an STM/AFM tip is coated with the same material as the sample under investigation[28]: the material is transferred during 'tip preparation' procedures such as voltage pulses. In fact, it is difficult to avoid this from happening. When scanning water-exposed oxide surfaces, an OH-terminated tip was reported to be the most stable and common termination, and the magnitude of the forces in Fig. 2 is comparable to previous nc-AFM measurements on $TiO_2$ with OH-terminated tips[28,29]. The quantitative agreement between the measured and calculated force curves in Fig. 2 is a strong indication that the calculated structure correctly reproduces the properties of the tip used in experiment.

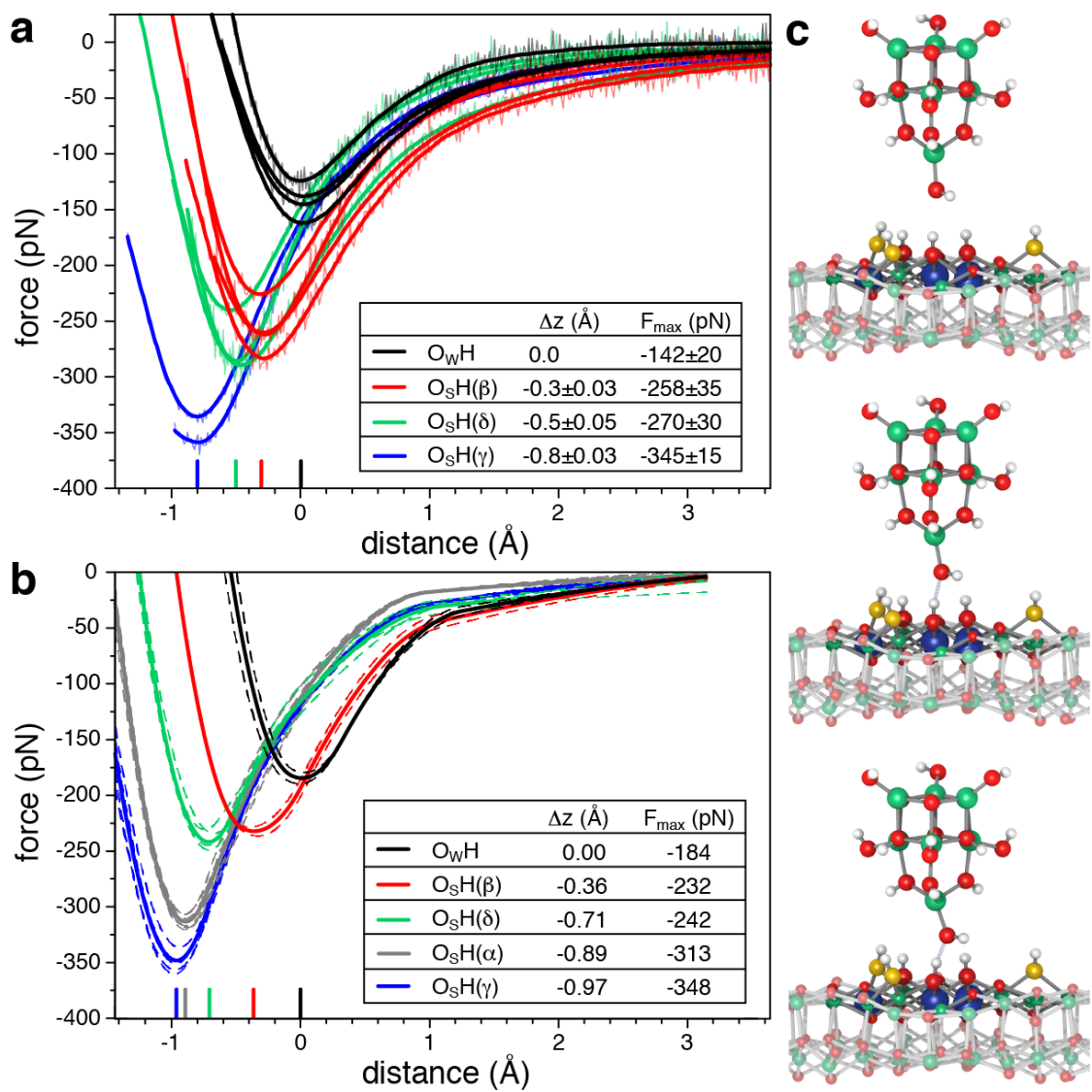

**Fig. 2 Probing individual surface hydroxyls with the AFM tip.** (a) Experimental short-range force-distance ($F(z)$) curves on the OH groups initially formed upon water dissociation, $O_WH$ and $O_SH(\beta)$, as well as on $O_SH(\delta)$ and $O_SH(\gamma)$ that were constructed by tip manipulation; four data sets are shown. (b) Calculated short-range force-distance curves. (c) Tip-sample configuration for various separations while probing an $O_SH(\beta)$, for full movies see the ExtendedData.

Note that the force curves were taken in the attractive regime, i.e., more negative values of $F$ imply that the tip experiences a stronger attraction to the surface. The panels (for full movies see the Supplementary Information) in Fig. 2c illustrate what happens: During the approach, a hydrogen bond forms between the O at the tip and the proton of the surface hydroxyl. The integration of the *F(z)* curves gives the interaction potential *E(z)*; the more attractive *F(z)*, the deeper the energy minimum of *E(z)*. Typically[30] the minima of *F(z)* and *E(z)* show a linear correlation (see also ExtendedData Fig. 11); thus, the force minimum of the *F(z)* curves is a direct measure of the strength of the hydrogen bond between the tip and the surface OH.

A surface hydroxyl will form weak H-bonds to the tip if its O–H covalent bond is strong (i.e., the surface O has a high PA). Conversely, weakly bound protons (located at surface O with a low PA) can form stronger H-bonds to the tip. The O at the tip and the O at the surface compete for the proton, hence the force minima of the *F(z)* curves serve as a direct measure of the proton affinity of the surface O of the probed OH group. Indeed, the values of the force minima in Fig. 2 follow the trend expected from the reactivity of the different sites; the tip experiences the highest attractive force from the proton bound to the least reactive surface site (lowest PA and strongest acid), $O_SH(\gamma)$. The H-bond between tip and surface becomes progressively weaker (less pronounced force minimum) when going from $O_SH(\gamma)$ to $O_SH(\delta)$ to $O_SH(\beta)$; the $O_WH$ exhibits the smallest interaction with the O atom of the tip (highest PA, weakest acid).

To further test how *F(z)* measurements relate to PA, and to put our results on a quantitative footing, we performed additional DFT calculations. We chose a series of probe molecules with a wide range of known PAs (see ExtendedData Tab. 1). Keeping the same tip model, we calculated *F(z)* curves on these molecules. In the calculations the molecules were oriented such that their OH group point towards the tip, similar to the OH groups on the $In_2O_3$ surface. The backbone of the molecules was fixed, and only the OH was allowed to relax in response to the approaching AFM tip.

Figure 3 shows the calculated minima of the *F(z)* curves as function of the PA of the probe molecules: the two quantities follow an almost perfect linear relationship. We use this scaling relation to derive values for the PAs of the different surface O atoms of the $In_2O_3$ surface from the measured *F(z)* curves (see shaded areas in Fig. 3). As an independent test of the reliability of this procedure, we compare the predicted ΔPAs of the surface sites with DFT calculations for the same quantity (grey arrows in Fig. 3); i.e., the change in adsorption energy when the proton is moved to various sites. We find good agreement.

Based on these results we propose to use AFM-measured *F(z)* curves as a direct measure of the PA of different surface O species; with quantitative determination of the proton affinity based on the relation in Fig. 3. Both, the mean value of PA of the surface hydroxyls on $In_2O_3$ and their variation compares well to the PA values assigned to hydroxyls in zeolites[19] where a large spread in PAs was attributed to a high reactivity in acid-base reactions. Knowledge about PA of individual OH surface groups is of direct importance for understanding the efficiency of de-/hydrogenation reactions on oxide surfaces in heterogeneous catalysis[31]. Here, particularly, $In_2O_3$ is of high interest as a promising hydrogenation catalyst for $CO_2$ conversion to methanol[22] and for the selective hydrogenation of acetylene[23].

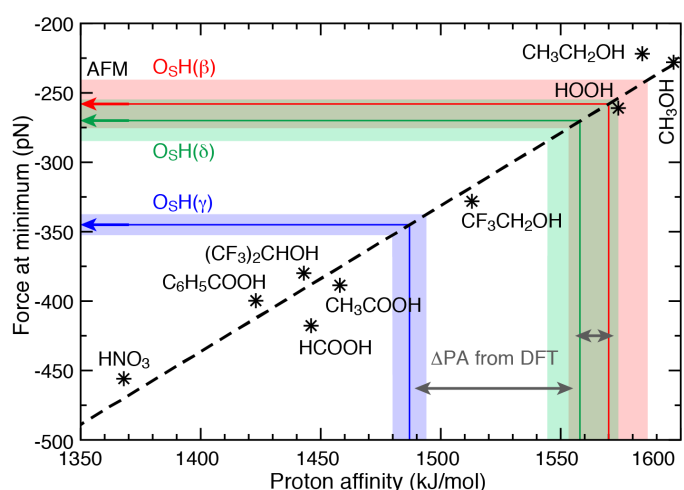

**Fig. 3 Calibration line relating AFM-measured force minima to proton affinity.** The minima in calculated *F-z* curves on various probe molecules, using the tip model shown in Fig. 3c, are related to their gas-phase proton affinities (PA). The experimental force minima for $O_SH(β)$, $O_SH(δ)$ and $O_SH(γ)$ are indicated by the areas in shaded in red, green, and blue, respectively. The differences, ΔPA, calculated by translating protons to the various $O_S$ positions, are shown by arrows.

The PA also governs the $pK_a$ in wet, solution-based chemical processes: an OH group with a strongly bound proton (high PA of the O atom) is a weak acid, and an OH group with a weakly bound proton (low PA of the O atom) is a strong acid. In addition to the PA, the $pK_a$ also includes the Gibbs free energy of solvation of the acid (XOH), the conjugated base (XO⁻), and the H⁺. These species have different solubilities, which leads to some scatter when we plot the $pK_a$ of our probe molecules as function of their PA (see ExtendedData Fig. 10). However, for the case of locally different OH groups on the same oxide surface we can assume that the solvation Gibbs free energies do not depend much on the specific site of the surface OH group. In this case we can provide an estimate for $pK_a$ differences between the surface OH groups from our determined PA values: using the slope of the $pK_a$ vs. PA graph from our probe molecules (see ExtendedData Fig. 10), a ΔPA between O(β) and O(γ) of about 83 kJ/mol translates to a p$K_a$ difference between O$_S$H(β) and O$_S$H(γ) of 5.2 p$K_a$ units, which is reasonable.

**Methods**

**Experiments.** The experiments were carried out in a two-vessel ultrahigh vacuum (UHV) system with base pressures of <2×10$^{-10}$ mbar and <2×10$^{-11}$ mbar in the preparation and analysis chamber, respectively. The preparation chamber contains standard surface cleaning and heating facilities. The analysis chamber is equipped with a low-temperature STM/AFM (Omicron) with a qPlus sensor[10] and a low-noise amplifier[32]. Tips were electrochemically etched from a 25 μm thick tungsten wire. All measurements were performed at a sample temperature of 5 K. The images and F(z) curves were recorded with different sensors: (1) $f_R$ = 31 kHz, $k$ = 2000 N/m, $Q$ ≈ 36,000, (2) $f_R$ = 47.5 kHz, $k$ = 3750 N/m, $Q$ ≈ 15,000 (3) $f_R$ = 77.7 kHz, $k$ = 5400 N/m, $Q$ ≈ 71,000. In constant-height nc-AFM typical oscillation amplitudes of 80-100 pm were used and a scanning speed corresponding to ~10 min per image. The tip was frequently refreshed on a Cu(100) single crystal and on the In$_2$O$_3$ sample using gentle voltage pulses until a frequency shift between 0 and -5 Hz in constant-current STM was obtained (setpoint +1 V, 20 pA).

In$_2$O$_3$(111) single crystals[33] were prepared by cycles of sputtering (1 keV Ar$^+$) and annealing at 450-500 °C in 6×10$^{-7}$ mbar oxygen as described in Refs.[25,26]. The clean surface was characterized with STM and AFM prior to the water adsorption experiments. Water was dosed via a high-precision leak valve from a glass tube containing a few ml of distilled water (Milli-Q). Prior to the experiments the water was cleaned by several freeze-pump-thaw cycles and the cleanliness was checked with a mass spectrometer.

The O$_S$H were manipulated with the STM/AFM tip. This causes the H atom to desorb from the surface as described in Ref.[27], or, though much less probable, to re-adsorb on a nearby surface O(3c) thereby forming a new O$_S$H group. The most successful way was voltage pulses during conventional STM imaging. The window for re-adsorption is approximately 0.1 V below the parameters for complete desorption (which are +2.8 V at an STM setpoint of 30 pA and +1.0 V) and slightly tip-dependent; examples are provided in the ExtendedData Fig. 5.

The force-distance curves were acquired by imaging in constant-height AFM mode and pausing the tip at a z-position close to the force minimum of the O$_W$H. The tip was then retracted by ~1.2 nm to cover a sufficiently wide range of the long-range interaction (the long-range interaction forces subtracted from the curves as described in ExtendedData Fig. 6). Conversion of the frequency-shift signal to forces was performed using Ref.[34]. The long-range forces[14] were treated by subtracting averaged background spectra measured in the same experiment on the $In_2O_3$(111) surface in-between the OH groups, i.e., mostly in regions A and C of the unit cell (see Fig. 1a). All displayed curves are averaged spectra of multiple repetitions on the same OH group or, for $O_WH$ and $O_SH$, also on different ones.

**Theory.** Density-functional theory (DFT) calculations were carried out with the plane-wave code PWscf of the Quantum Espresso software package[35] using the Perdew-Burke-Ernzerhof PBE exchange-correlation functional[36], Vanderbilt ultrasoft pseudopotentials[37], and a plane wave kinetic energy cutoff of 30 Ry. The In-4$d$ electrons were treated as valence states. The same setup was used in our previous work, see Refs.[25-27]. The van der Waals forces between the AFM tip and the substrate were included by using the Grimme D3 dispersion correction with Becke-Johnson damping[38]. The $In_2O_3$(111) surface structures were represented by a symmetric slab with a thickness of four $O_{12}$–$In_{16}$–$O_{12}$ tri-layers. The calculated theoretical bulk lattice constant of 10.276 Å was used for the in-plane dimensions of the periodically repeated slabs, which were separated by an 18 Å vacuum region. In the AFM calculations the thickness of the vacuum region was increased to 36 Å to accommodate the tip.

The adsorption of water and the displacement of protons were studied first with a primitive hexagonal (1×1) surface unit cell (160 atoms without adsorbates, lateral dimension: 14.532 Å). In this cell a displaced proton always stays next to the water triangle/propeller where it originates from. To study also the situation where a proton is displaced to a neighboring, intact water triangle/propeller, calculations were repeated for a three times larger hexagonal (√3×√3) unit cell (480 atoms, lateral dimension: 25.171 Å). For the smaller cell a (2,2,1) Monkhorst-Pack $k$-point mesh was used, whereas the Γ–point was sufficient for the larger cell. An initial set of AFM calculations was done with the (1×1) unit cell. All final results, however, were obtained for the larger (√3×√3) cell. In the geometry optimizations the bottom two tri-layers of the slab were kept fixed. Only the atoms in the upper two tri-layers together with the adsorbed water molecules were relaxed, using a force convergence threshold of 5 meV/Å. The adsorption energy of H atoms was calculated with respect to the $H_2$ molecule, i.e., $E_{ads}^{H} = E_{slab}^{ref} + \frac{1}{2}E_{mol}^{H_2} - E_{slab}(H)$ .

The proton affinity $\Delta H_{PA}$ of a surface site O is defined as the negative enthalpy change of the protonation reaction in the gas phase

$$X\text{-}O^-_{(g)} + H^+_{(g)} \rightarrow X\text{-}OH_{(g)} \ .$$

The closely related gas-phase basicity $\Delta G_{GPB}$ is the negative of the Gibbs free-energy change for the same gas-phase reaction. In contrast, the acidity constant p$K$a is derived from the Gibbs free-energy change of protonation $\Delta G_{acid}$ in a liquid environment. $\Delta G_{acid}$ differs from the gas-phase basicity by including the solvation free energies $\Delta G_{sol}$ of all participating species and can be derived by a thermodynamic cycle

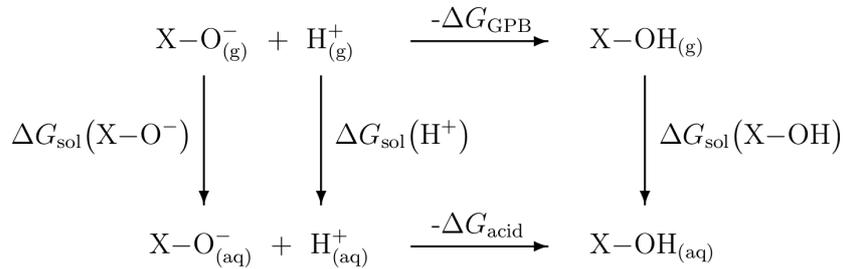

The acidity constant p$K_a$ is the logarithm of the corresponding equilibrium constant $K$

$$K = e^{\Delta G_{acid}/kT} \quad \text{and} \quad pK_a = \log K \ .$$

Note the small energy scale: at room temperature a change of only 6 kJ/mol in the Gibbs free energy $\Delta G_{acid}$ leads to a shift of the p$K_a$ value by one unit.

An extensive search was performed to find a realistic atomistic model for the AFM tip. Several tip models with different structure and composition were considered. Many tip models were unstable in initial test calculations of $F(z)$ curves: some of them lost their terminal OH group, or protons were transferred from the surface to the tip (converting the terminal OH group into a water molecule). For a stable tip, which is able to describe the experimentally measured $F(z)$ curves, it is mandatory to find a structure that is not too reactive. Good criteria for a chemically inert tip are high coordination numbers of the atoms and a large HOMO-LUMO gap. The final AFM tip used in all calculations is based on a water-saturated indium oxide cluster. Its structure is described in ExtendedData Fig. 8.

In the calculations of the force--distance curves $F(z)$ we followed closely the procedures of the AFM experiments. The tip was initially placed at a vertical position $z$ close to the force minimum, with its center above the O atom of a particular OH group (see choice of $z$=0 in the AFM experiments). The bottom two tri-layers of the slab and the top atoms of the tip were kept fixed (all atoms except the tip In apex atom together with its four OH groups, see ExtendedData Fig. 8c); all other atoms were allowed to relax. Starting from the initial height, the $z$-coordinate of the AFM tip was increased and decreased in steps of 0.1 Å. The atomic positions of the previous step were used as initial configuration in the geometry optimization of the subsequent step. At larger $z$-distances the step size was increased to 0.24 Å. This procedure was typically repeated for 50 $z$-values. The result of the whole sequence is shown for two examples (the $O_WH$ and $O_SH(\beta)$ groups and one tip orientation) in the Supplementary Movies. At each tip height $z$ the energy $E(z)$ of the optimized geometry was recorded. The $F(z)$ curves were then obtained by fitting a spline to the resulting $E(z)$ curves and taking the derivative. Jumps in the $E(z)$ curves were frequently observed, which were caused by the re-orientation of the tip OH group or an OH at the surface. The new local energy minimum structure was then used as starting point for the calculation of a new $E(z)$ curve following the iterative procedure described above.

The background correction of the calculated $F(z)$ curves was done in the same way as in experiment. Additional $F(z)$ curves were calculated above the high-symmetry sites A and C at the hydroxylated and the water-free $In_2O_3(111)$ surface using two different azimutal orientations of the tip (see ExtendedData Fig. 9). On the hydroxylated surface, the A and C sites are as far away as possible from the adsorbed water molecules. The curves were averaged and subtracted from all $F(z)$ curves calculated above surface OH groups (see Fig. 2b in the manuscript).

For the calculation of force-distance curves for $O_SH(\alpha)$, $O_SH(\gamma)$ and $O_SH(\delta)$ groups a proton was displaced on the hydroxylated $In_2O_3(111)$ surface from $O_SH(\beta)$ to the respective surface site. A (1×1) surface unit cell contains three O atoms of each type $\alpha$, $\gamma$ and $\delta$. Thus, 9 different $O_SH$ groups can be formed, which are all symmetry-inequivalent as the displacement of the proton breaks the 3-fold symmetry of the initial triangle/propeller-like water structure. Calculations were performed for all 9 $O_SH$ groups. In addition, for each surface OH (including $O_SH(\beta)$ and $O_WH$), 6 different orientations of the AFM tip were considered, with the terminating OH group of the tip pointing in different azimuthal directions, rotated by 60°. Some of the calculated $F(z)$ curves turned out to be identical since the OH group of the tip rotated into the same direction in the geometry optimizations. A selected set of the final distinctly different 25 $F(z)$ curves is shown in Fig. 2b of the manuscript. The scatter in the curves demonstrates the uncertainty due to specific surface site and tip orientation. Each of the $F(z)$ curves can be assigned unambiguously to one of three classes corresponding to the three types of surface OH groups $O_SH(\alpha)$, $O_SH(\gamma)$ and $O_SH(\delta)$.

Curves within one class were averaged and are discussed in the main manuscript. The three-times larger (√3×√3) surface unit cell offers 27 different sites for a displaced proton from an $O_SH(β)$ group, 9 of each type α, γ and δ. Here, force-distance curves were calculated only for a selected set of possible $O_SH$ groups. The selection was based on the results from the (1×1) unit cell and on the observed jumps in the AFM experiments.

For each probe molecules an *E(z)* curve was calculated using the same AFM tip as in our surface calculations and keeping the geometries, the overall procedure, and all calculation parameters as similar as possible. The same supercell was used, and the probe molecules were oriented such that their OH group pointed toward the AFM tip. Again, as in the surface calculations, different orientations of the AFM tip with respect to the OH-axis of the molecule were probed.

**Acknowledgements**

This work was supported by the Austrian Science Fund (FWF), project T 749-N27 (Herta-Firnberg-Stelle, MW) and Z 250-N27 (Wittgenstein Prize, UD), as well as the German Research Foundation (DFG), Research Unit FOR 1878 (funCOS, BM). MW and UD also acknowledge funding under the Horizon 2020 Research and Innovation Programme under the Grant Agreement No 810626.

**Contributions**

MW and MS conducted the experiments, MW, MS, and MSch analyzed the data, BM conducted the calculations, MW, BM and UD wrote the manuscript, which was reviewed and edited by all authors, UD oversaw the project.

The authors declare no competing interests.

Supplementary Information is available for this paper.